\documentclass{PoS}

\title{Quasar host galaxies in the SDSS Stripe 82}

\ShortTitle{Quasar host galaxies in the SDSS Stripe 82}

\author{\speaker{Jari Kotilainen}\\
Finnish Centre for Astronomy with ESO (FINCA), University of Turku, Finland\\
        E-mail: \email{jarkot@utu.fi}}

\author{Renato Falomo\\
        INAF -- Osservatorio Astronomico di Padova, Italy\\
        E-mail: \email{renato.falomo@oapd.inaf.it}
}

\author{Daniela Bettoni\\
        INAF -- Osservatorio Astronomico di Padova, Italy\\
        E-mail: \email{daniela.bettoni@oapd.inaf.it}
}

\author{Kalle Karhunen\\
        Tuorla Observatory, Department of Physics and Astronomy, University of Turku, Finland\\
        E-mail: \email{kajukar@utu.fi}
}

\author{Michela Uslenghi\\
        INAF - IASF, Milano, Italy\\
        E-mail: \email{uslenghi@lambrate.inaf.it}
}

\abstract{We present first results from our study of the properties of\ 
$\sim$400 low redshift (z $<$ 0.5) quasars, based on a large homogeneous\
 dataset derived from the Stripe 82 area of the Sloan Digital Sky Survey\ 
(SDSS) Data Release 7 (DR7). For this sky region, deep (r$\sim$22.4) u,g,r,i,z\
images are available, up to $\sim$2 mag deeper than standard SDSS images,\ 
allowing us to study both the host galaxies and the Mpc-scale environments of\ 
the quasars. This sample greatly outnumbers previous studies of low redshift\ 
quasar hosts, from the ground or from space. Here we report the preliminary\ 
results for the quasar host galaxies. We are able to resolve the host galaxy\ 
in $\sim$80 \% of the quasars. The quasar hosts are luminous and large, the\ 
majority of them in the range between M*-1 and M*-2, and with $\sim$10 kpc\ 
galaxy scale-lengths. Almost half of the host galaxies are best fit with an\ 
exponential disk, while the rest are spheroid-dominated. There is a reasonable\
 relation between the central black hole mass and the host galaxy luminosity.}

\FullConference{Nuclei of Seyfert galaxies and QSOs - Central engine \& conditions of star formation\\
                 November 6-8, 2012\\
                 Max-Planck-Insitut f\"ur Radioastronomie (MPIfR), Bonn, 
Germany}

\begin{document}

\section{Introduction}

Supermassive black holes (SMBHs) are ubiquitously found in the centres of 
massive galaxies and the BH masses correlate with the large-scale properties 
of the galaxies, e.g., the stellar velocity dispersion, the luminosity, and 
the mass of the spheroidal component (e.g. Gultekin et al. 2009). 
These relations have been interpreted as the outcome of a joint evolution 
between BHs and their host galaxies and are therefore of critical importance 
for understanding the processes that link nuclear activity to galaxy formation 
and evolution 
(e.g. Decarli et al. 2010; Merloni et al. 2010; Cisternas et al. 2011). 

Nuclear activity, i.e. accretion onto the SMBH, appears to be a common phase 
in the evolution of normal galaxies. Furthermore, SMBHs may well have a period 
of maximum growth (maximum nuclear luminosity) contemporaneous with the bulk 
of the initial star formation in the bulge. 

Both ground-based and HST studies have established that practically all 
luminous low redshift (z$<$0.5) quasars reside in massive, spheroid-dominated 
host galaxies, whereas at lower luminosities quasars can also be found also in 
early-type spiral hosts at (e.g. Bahcall et al. 1997; Dunlop et al. 2003; 
Pagani et al. 2003; Floyd et al. 2004; Jahnke et al. 2004). 
This is in good agreement with the BH -- bulge relationship in inactive 
galaxies (e.g. Gultekin et al. 2009), 
since very massive BHs power luminous quasars. 
Only $\sim$15\% are mergers but it is difficult to determine clear merger 
signatures from morphology alone. 

At high redshift (z $>$1) HST observations of quasar host galaxies 
(e.g. Floyd et al. 2013 and references therein) have been complemented by 
significant contributions from 8-m class ground- based telescopes under superb 
seeing conditions (Kotilainen et al. 2007, 2009) or with adaptive optics
(Falomo et al. 2005, 2008; Wang et al 2013).

Comparison of host galaxies of AGN at high and low redshift constrain host 
galaxy evolution, in comparison to the evolution of normal galaxies. 
Most of the quoted studies of quasar hosts have considered only few tens of 
objects, therefore in order to derive a picture of the host properties at 
various redshifts one needs to combine many different, often heterogeneous, 
samples. Observations carried out by the HST are more homogeneous than samples 
based on ground-based observations but the size of these samples remains 
relatively small. For instance in the range 0.25 $<$ z $<$ 0.5 only about 
50 quasar hosts have been imaged by HST.

In order to explore a significantly larger dataset of quasars, one should 
refer to large surveys that include both imaging and spectroscopic data. 
In this respect one of the most productive recent surveys is 
the Sloan Digital Sky Survey (SDSS).  Standard SDSS images are, however, 
too shallow and the faint nebulosity around the nucleus of quasars is not 
easily detected. This problem has been overcome in the case of a special sky 
region mapped by the SDSS called the Stripe82 (Annis et al. 2011). It covers 
a total area of $\sim$270 sq.deg that was observed $\sim$80 times, and 
the co-added images are up to $\sim$2 mag deeper than the standard SDSS images. 

Here we present preliminary results from our study of the properties of 
the host galaxies of the largest so far, homogeneous sample of low redshift 
(z $<$ 0.5) quasars in the Stripe 82 area of the SDSS DR7. Full results on 
the host galaxies will be discussed in Falomo et al. (in prep.) and on the 
environments in Karhunen et al. (in prep.) We adopt the concordance cosmology 
with H$_0$ = 70 km s$^{-1}$ Mpc$^{-1}$, 
$\Omega_m$ = 0.3 and $\Omega_\Lambda$ = 0.7.

\section{The SDSS Stripe 82 quasar sample}

To derive the sample of low redshift quasars, we used the fifth release of 
the SDSS Quasar Catalog (Schneider et al. 2010) that is based on the SDSS DR7. 
It consists of quasars that have a highly reliable redshift measurement, 
$i\leq15.0$, absolute magnitude $M_i<$-21.0, and at least one emission line 
with FWHM$>$1000 km/sec. This catalog contain $\sim$10$^5$ spectroscopically 
confirmed quasars. Our analysis is done for low redshift objects that are 
in the Stripe82 region, thus making possible the study of the  host galaxies.


The final sample consists of 416 quasars at 0.1 $<$ z $<$ 0.5. 
The sample is dominated by radio-quiet quasars: only 24 (5\%) are radio-loud. 
The mean redshift of the sample is $<z>$ =$0.39\pm0.08$ and the mean 
absolute magnitude is $<M_i>$ = $-22.68\pm0.61$. 
We also selected a comparison sample of 580 inactive galaxies matched closely 
in redshift with the quasars, that will be used in a later stage of this 
project to compare the properties of the environments.

\section{Image analysis}

In order to derive the properties of the quasar host galaxies, we performed 
a 2D fit of the images of the quasars, assuming the superposition of two 
components: the nucleus and the surrounding nebulosity. The first is described 
by the local Point Spread Function (PSF) of the image, while the second is 
described by a galaxy model following a Sersic law convolved with the PSF. 
The analysis was performed using the Astronomical Image Decomposition Analysis 
(AIDA, Uslenghi \& Falomo 2011).

The most critical aspect of the decomposition is the determination of the PSF. 
The field of view of SDSS images is large enough to always contain many stars 
to derive the most suitable PSF. 
In each field, we have selected a number of stars (between 5 and 15) around 
the target, based on various parameters such as their magnitude, FWHM and 
ellipticity. 
We then defined a radius to compute the PSF model and a ring around each star 
where to compute the sky background. 
The PSF model was then obtained from the fit of all selected stars using 
a multifunction 2D model composed of three gaussians and one exponential 
function.

\begin{figure}[h]
 \includegraphics[width=14truecm]{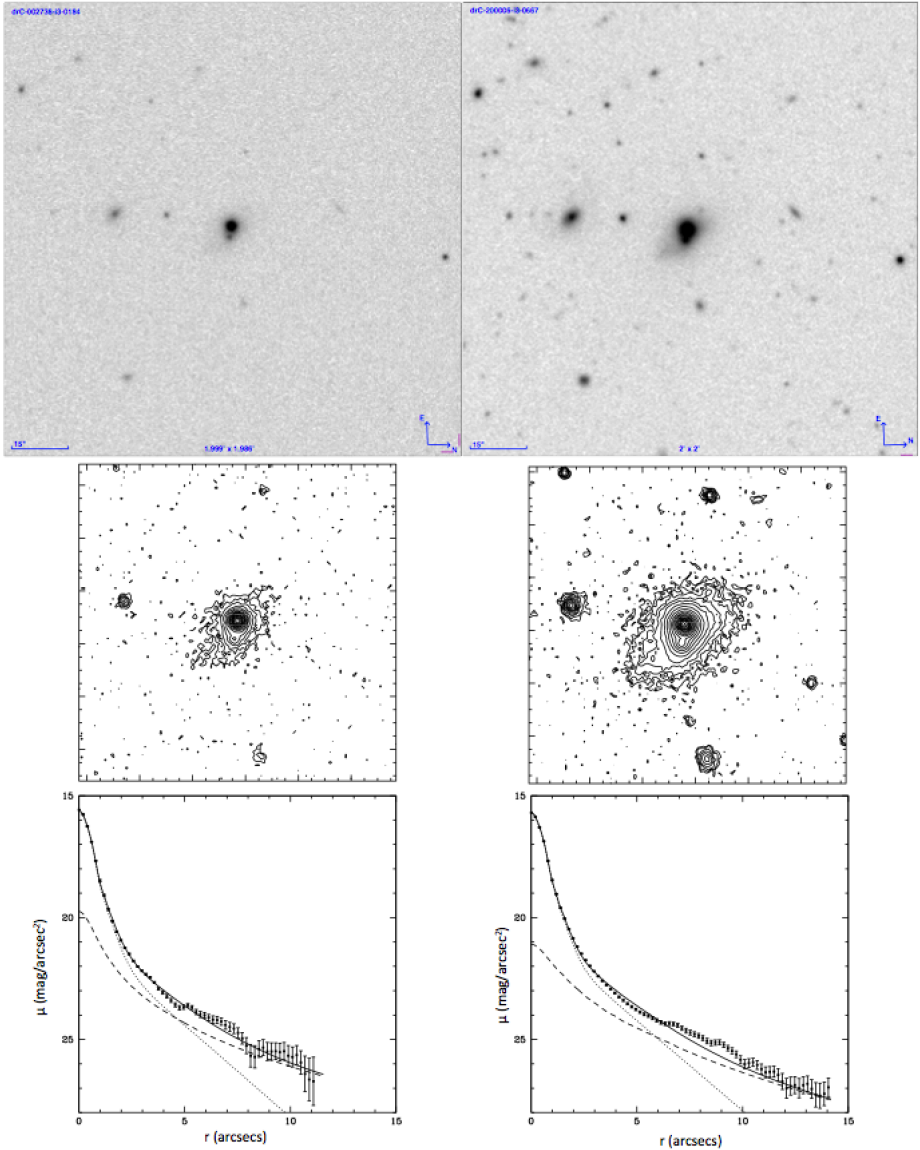} 
 \caption{Example of the analysis. Top: the SDSS DR7 i-band image 
(left), and the Stripe 82 i-band image (right), highlighting the gain in depth. 
Middle: the isophotes of the central region. 
Bottom: the luminosity profiles with the best-fit (solid line) consisting of 
the nucleus (PSF; dotted line) and the host galaxy convolved with the PSF 
(Sersic model; dashed line). The host galaxy is well resolved. }
   \label{analysis}
\end{figure}

The next step of the analysis is to fit each quasar with both a scaled PSF and 
a two-component model (PSF + galaxy). 
In order to distinguish resolved from unresolved objects, we compared 
the $\chi^2$ of the two fits and visually inspected all the fits. 
In Figure \ref{analysis} we show an example of the adopted procedure. 

\section{Quasar host galaxies}

From the above procedure, we are able to classify 354 (85\%) quasars as 
resolved, and 56 (13\%) quasars as unresolved. Further 6 (1\%) quasars were 
contaminated by nearby bright sources or defects, and were removed from 
following analysis.  
Of all the quasars in the sample 155 (37\%) have a nearby companion or tidal 
features. 

The average K-corrected rest-frame R-band absolute magnitude of 
the host galaxies for all the resolved quasars is $<M(R)>$ = -22.96 $\pm$ 0.62. 
This is in excellent agreement with a smaller sample of quasar hosts in the 
same redshift range observed by HST (Floyd et al. 2004), 
$<M(R)>$ = -23.00 $\pm$ 1.05. 

\begin{figure}[h]
 \includegraphics[width=8truecm]{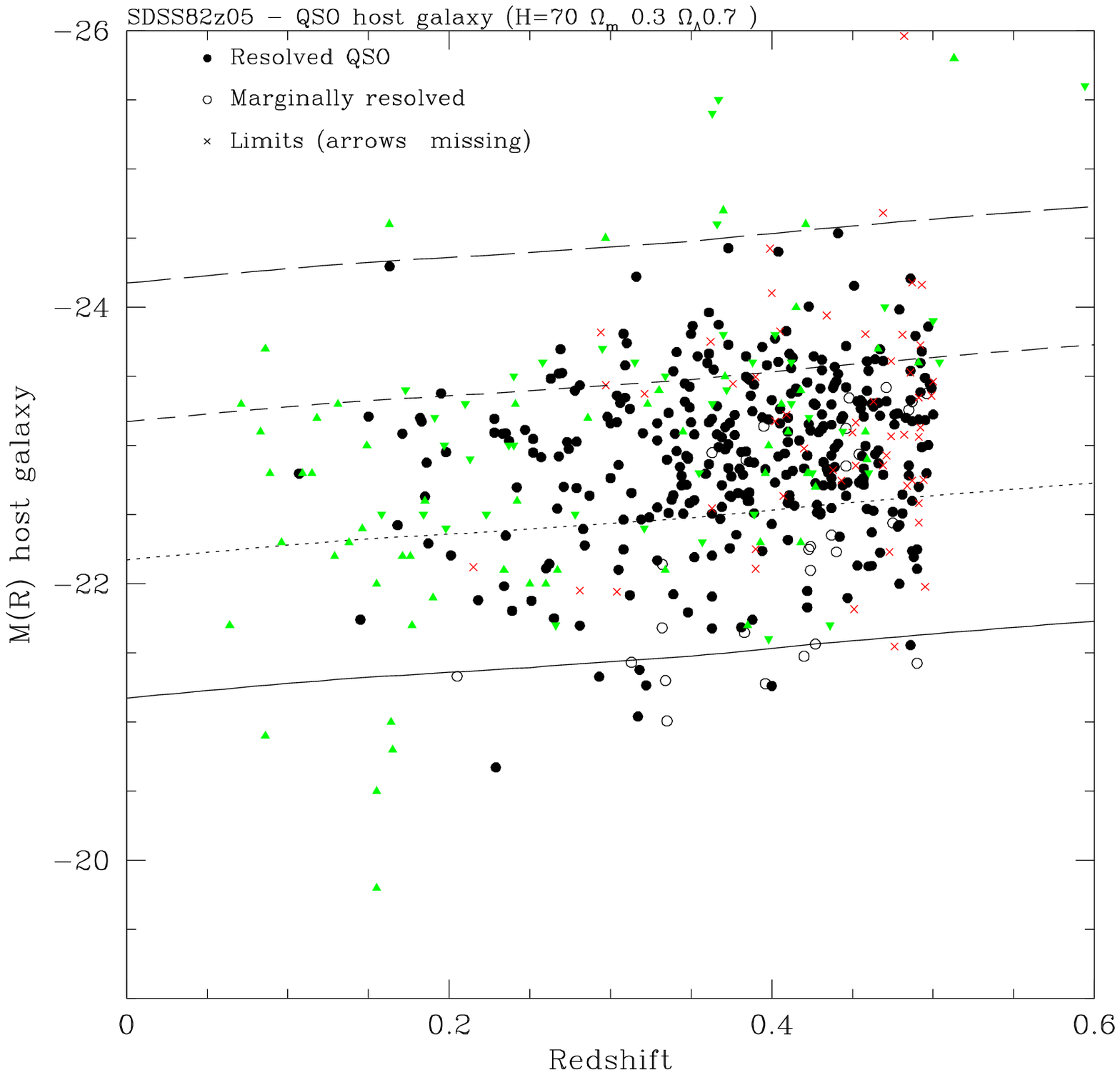}
 \includegraphics[trim=0 20 0 0, clip,width=8truecm]{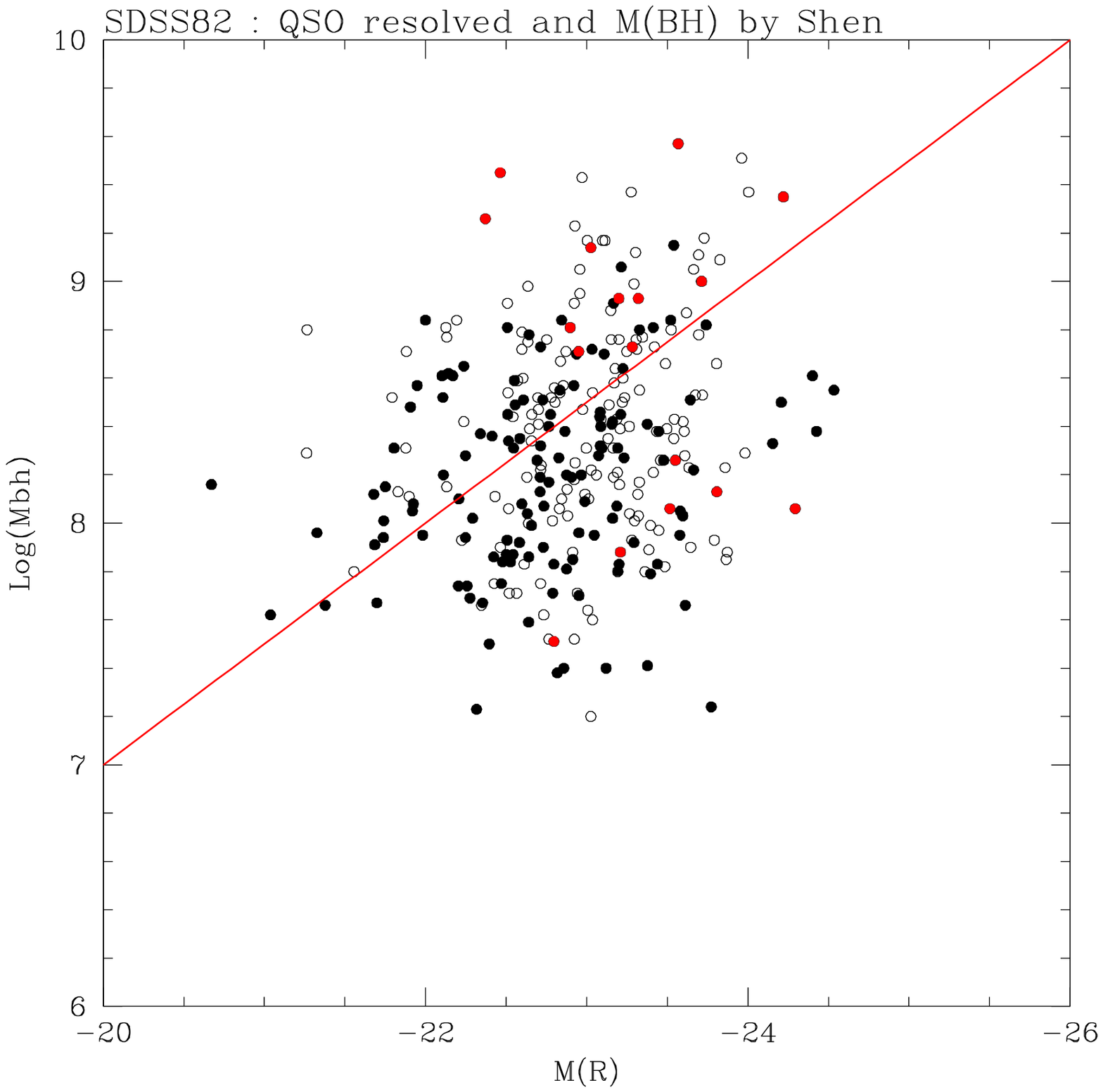}
 \caption{
Left: The absolute magnitude of quasar host galaxies versus redshift. 
Resolved quasars (filled circles), marginally resolved (open circles) and 
luminosity lower limits (red crosses with arrows) . For comparison we include 
a compilation of $\sim$100 quasar host galaxies from HST observations by 
Decarli et al 2010 (filled green triangles: inverted triangles for 
radio-loud objects).
Right: Absolute magnitude of quasar host galaxies versus BH mass for resolved 
quasars. The solid line is the Bettoni et al. (2003) relation.
Open points are quasars with poor spectra and uncertain BH masses. 
Red points are radio-loud quasars.}
   \label{host}
\end{figure}

The distribution of the quasar host galaxies in the redshift-luminosity plane 
(Fig \ref{host}, left) shows that they are encompassed between M(R) $\sim$-21 
and M(R) $\sim$-24, corresponding to a range between M* and M*-3 with 
the majority of them in the range between M*-1 and M*-2. 

The host galaxy sizes of the resolved quasars range from $\sim$2 kpc to 
$\sim$15 kpc, with average value $\sim$8 kpc. This is in good agreement with 
previous studies (Dunlop et al. 2003, Bennert et al. 2008) and indicates that 
quasars preferentially live in large host galaxies. Both spheroid- and 
disc-dominated morphologies were found in our sample, with almost half of 
the sample best fit with n=1, i.e. an exponential disk profile. Work is in 
progress to determine if there is a dependence of the host galaxy morphology 
(Sersic index) with the luminosity of the quasar, as found in previous studies
(e.g., Dunlop et al. 2003).

\section{Black hole -- galaxy relationship}

Our large and homogeneous dataset allows us to investigate the relationship 
between the BH mass and host galaxy mass for low redshift quasars. For the BH 
mass, we adopted the measurements obtained by Shen et al. (2011) who estimate 
the virial BH mass using the FWHM of $H_{\beta}$ for all the quasar with 
$z<0.7$. All the spectra have been visually inspected and 31 objects 
($\sim$7\%) have been removed from the final sample because of very low 
S/N ratio of the spectra. For three objects we have obtained a new measurement 
of the BH mass. 

In Figure \ref{host}, right, we show the relationship between the BH mass and 
the absolute magnitude of the host galaxy in the R band for all the resolved 
quasars that have good S/N spectra. The absolute magnitude of the host 
galaxies is in the range -22 $<$ M(R) $<$ -24 and BH masses between 
10$^7$ and 10$^9$ M$\odot$. The average BH mass of the quasars is 
$<$log(MBH)$>$ = 8.33$\pm$0.43. The relation between the BH masses and 
the host galaxy luminosities (masses), albeit on average agrees with 
the relation in the local Universe, exhibits  quite large scatter. 

\section{Future work}

For the $\sim$100 well resolved quasars at z $<$ 0.25 and m(host) $<$ 21, 
we shall investigate the host galaxy colours in the bgr rest-frame. These will 
have implications on the fraction of blue star forming host galaxies as a 
function of quasar type, luminosity and environment (e.g. Jahnke et al. 2004; 
Floyd et al. 2013). We are also investigating the properties of the 
close ($<$ 0.5 Mpc) environment of the quasars.

\end{document}